\documentclass[journal]{IEEEtran}
\usepackage{graphicx}
\usepackage{epstopdf}
\begin{document}
\title{Calibration of Gamma-ray Burst Polarimeter POLAR
\author{H.L. Xiao, W. Hajdas, T.W. Bao, T. Batsch, T. Bernasconi, I. Cernuda,  J.Y. Chai, Y.W. Dong, N. Gauvin, M. Kole, M.N. Kong, S.W. Kong, 
L. Li, J.T. Liu, X. Liu,  R. Marcinkowski,  S. Orsi, M. Pohl, N. Produit, D. Rapin,
A. Rutczynska, D. Rybka, H.L. Shi, L.M. Song, J.C. Sun, J. Szabelski, B.B. Wu, R.J. Wang, X. Wen, H.H. Xu, L. Zhang,   L.Y. Zhang, S.N. Zhang, X.F. Zhang, Y.J. Zhang, A. Zwolinska}
\thanks{Manuscript received December 7th, 2015.}
\thanks{H.L. Xiao, W. Hajdas and  R. Marcinkowski are with Paul Scherrer Institut, 5232 Villigen PSI, Switzerland (e-mail: hualin.xiao@psi.ch).
H.L. Xiao is also with Institute of High Energy Physics, Chinese Academy of Science, Beijing 100049, China.}
 \thanks{ T.W. Bao,  J.Y. Chai, Y.W. Dong, H.L. Shi, L.M. Song, J.C. Sun, B.B. Wu, R.J. Wang, X. Wen, H.H. Xu, L. Zhang,   L.Y. Zhang, S.N. Zhang, X.F. Zhang,  Y.J. Zhang, M.N. Kong, S.W. Kong,  L. Li, J.T. Liu and X. Liu
are with Key Laboratory of Particle Astrophysics, Institute of High Energy Physics, Chinese Academy of Sciences, Beijing 100049, China.}
 \thanks{M. Kole, S. Orsi, M. Pohl and D. Rapin are with University of Geneva (DPNC), quai Ernest-Ansermet 24, 1205 Geneva, Switzerland.}
\thanks{N. Produit,  I. Cernuda and N. Gauvin are with
 University of Geneva, ISDC Data center for Astrophysics, 16, Chemin d'Ecogia, 1290 Versoix Switzerland.}
 \thanks{A. Rutczynska, D. Rybka, A. Zwolinska  J. Szabelski, T. Batsch and T. Bernasconi are with  
 National Centre for Nuclear Research ul. A. Soltana 7, 05-400 Otwock, Swierk, Poland.}
 }

\newcommand{\vect}[1]{\mathbf{#1}}
\newcommand{\vecm}[1]{\mathbf{#1}}

\maketitle
\pagestyle{empty}
\thispagestyle{empty}
\begin{abstract}
Gamma Ray Bursts (GRBs) are the strongest explosions in the universe
which might be associated with creation of black holes.
Magnetic field structure and burst dynamics may influence polarization of the emitted gamma-rays.
Precise polarization detection can be an ultimate tool to unveil the true GRB mechanism.
POLAR is a space-borne Compton scattering detector for precise measurements of the GRB polarization.
It consists of a 40$\times$40 array of plastic scintillator bars
read out by 25 multi-anode PMTs (MaPMTs). It is scheduled to be launched into space in 2016 onboard of the Chinese space laboratory TG2.
We present a dedicated methodology for POLAR calibration and some calibration results based on the combined use of the laboratory radioactive
sources and polarized X-ray beams from the European Synchrotron Radiation Facility. 
They include calibration of the energy response, computation of the energy conversion factor vs. high voltage 
as well as determination of the threshold values, crosstalk contributions and polarization modulation factors.

\end{abstract}

\section{Introduction}

\IEEEPARstart{G}{amma}-ray bursts (GRBs) discovered more than 40 years ago, are short and intense flashes of gamma-rays produced at cosmological distances and appearing randomly  in the sky and in time.
They are the strongest explosions in the universe which might be associated with creation of black holes.
Since their discovery, thousands of  GRBs have been detected by various dedicated detectors.
Despite of it the emission mechanism of GRBs still remains a mystery.
Direct measurements of the polarization in the hard X-ray and gamma-ray band are thought to be capable of unveiling their emission mechanism \cite{zb2,zb1}.

POLAR is a novel, compact, space-borne instrument dedicated for measurements of the polarization 
in the prompt emission of the gamma-ray bursts.
It has both a large effective detection area ($\sim$ 80 cm$^2$) and a large field of view ($\sim$ 1/3 of full sky). 
The instrument was developed by an international collaboration of China, Switzerland and Poland.
Polarization measurements in POLAR use Compton scattering process and are based on detection of energy depositions 
in its 1600 scintillator bars.
POLAR is planned to be launched into space in 2016 on-board the Chinese space laboratory TG2.
Before the launch the basic instrument parameters such as energy response,  energy calibration factor vs. high voltage dependence, threshold values,
crosstalk and modulation factors had to be determined in series of calibration, test and verification campaigns.
In this paper we present a wide calibration methodology which combines the use of the laboratory radioactive sources and
polarized X-ray beams from the European Synchrotron Radiation Facility (ESRF).
\section{POLAR Instrument}

\begin{figure}[htb]
\centering
\begin{center}
\includegraphics[width=2in]{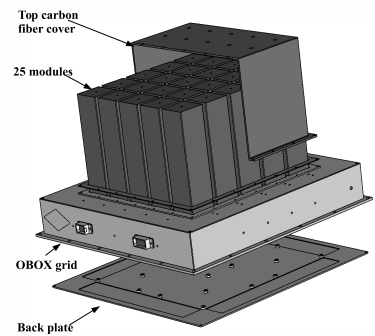}
\caption{Exploded view of full POLAR instrument. }
\label{polardet}
\end{center}
\end{figure}

\begin{figure}[htb]
\begin{center}
\includegraphics[width=1.8in]{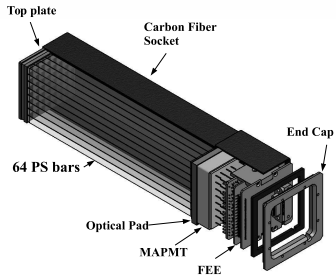}
\caption{Exploded view of POLAR detector module. }
\label{polarmodule}
\end{center}
\end{figure}

POLAR consists of 25 detector modules (see. Fig.~\ref{polardet}).
In total, they encompass 1600 plastic scintillator bars working as a POLAR gamma-ray target.
Each scintillator bar has dimension of 5.8 $\times$ 5.8 $\times$ 176 mm$^3$. It is wrapped in a highly reflective foil.
Each of 25 identical detector modules contains a matrix of with $8 \times 8$ bars.
The module is read-out by a multi anode photomultiplier (MaPMT, Hamamatsu H8500),
mechanically coupled to the bottom of the scintillator bars via a thin transparent optical pad. 
Each module is enclosed in a 1 mm thick carbon fiber socket (see Fig.~\ref{polarmodule}).
Signals from the MaPMT are firstly processed by ASIC (VA64PMT from IDEAS) based front-end electronics boards directly connected with the photomultipliers. 
After pre-processing the data is sent to the POLAR Center Task Processing Unit Unit for further analysis, merging and download.

The principle of polarization detection by POLAR is based on the fact that gamma-rays undergoing
Compton scattering in the plastic scintillator bars tend to scatter perpendicularly to their polarization direction
\cite{silvio,nicolas}.
The distribution of the azimuthal  angles of the photons that are Compton scattered between plastic scintillator bars of POLAR 
(called modulation curve) follows characteristic pattern. It has a sinusoidal shape for polarized gamma-rays or a
flat curve for unpolarized gamma-rays.
The amplitude of the sinusoidal curve is dependent on the polarization level.
Polarization degree of detected gamma-rays can be subsequently obtained by 
studying of the modulation curve features and comparing them with Monte Carlo simulations.

\section{Calibration with laboratory radioactive sources}
 \subsection{Energy conversion factor vs. PMT high voltage dependence}
\begin{figure}[!t]
\begin{center}
\includegraphics[width=2.5in]{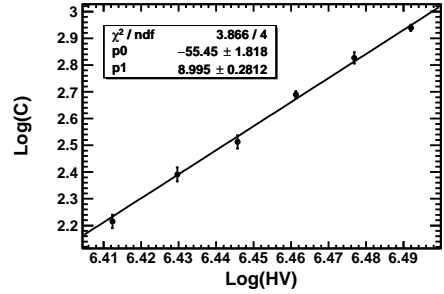}
\caption{An example of the dependence of the energy conversion factor  vs. the high voltage applied to the MaPMT. }
\label{hvvsce}
\end{center}
\end{figure}

 Energy conversion factor $C$ is defined as the recorded energy deposition (in units of ADC channel) per unit of real energy deposition (in units of keV). 
The gain of a PMT $G$ as a function of  the high voltage $V$
applied to the PMT can be given by $G=\alpha V^{\beta}$, where $\alpha$ and $\beta$ are two parameters. 
We assume that the energy response of our detector is linear in our interested energy range between 50 keV - 500 keV for the incoming GRB photons. 
The energy conversion factor $C$ as a function of high voltage can therefore be given by
\begin{equation}
C=A \cdot V^{\beta}, 
\label{eq:hvc}
\end{equation}
where $A$ is the model parameter. 
Measuring the energy conversion factor $C$ vs. high voltage dependence $V$ can provide the model parameters.
It should be noted that $A$ and $\beta$ vary from channel to channel because of the gain non-uniformity of MaPMT.
To obtained the dependences, we used the Cs-137 gamma-ray radioactive source placed in front of the FM. 
The data were taken at six high voltage values ranging from 610 V to 660 V with a step of 10 V. 
Energy conversion factors were obtained by subsequent fitting of the Compton edges in the recorded energy spectra (Compton edge energy value is equal to 470 keV).
Fig.~\ref{hvvsce} shows in the double logarithmic scale the values of the energy conversion factors for one readout channel vs. high voltage values. 
A linear fit of the logarithms of the data points provides the parameters of the gain vs high voltage dependence for this channel.
The relation between energy conversion factors and high voltage values allows for determination of any energy deposition at a given high voltage value and ADC channel.
This dependence can also be used to optimize both the dynamic range and the minimum photon energy (energy threshold)
 of the POLAR FM in order to obtain the maximum detection counting rate from the GRB and the best signal to background ratio.

\subsection{Threshold}
\label{sec:thr}

\begin{figure}[!t]
\begin{center}
\includegraphics[width=2.5in]{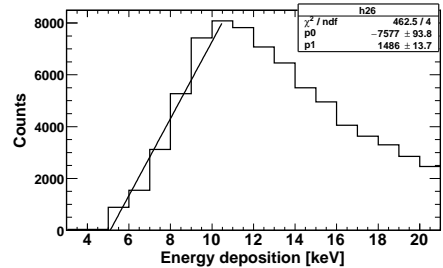}
\caption{An example of fit of threshold. }
\label{thresholdfit}
\end{center}
\end{figure}
Precise knowledge of all 1600 energy threshold values is crucial because the experimental modulation factors used for reconstruction of the GRB polarization degree are threshold dependent.
The ASICs in POLAR front-end electronics have a comparator for each channel. 
In addition to the energy depositions, the logical outputs from the comparators are also recorded and used to make  
a first trigger decision.
There is a sharp cut-off on the left side of the energy spectrum of the triggered hits (see Fig.~\ref{thresholdfit}).
Fitting it with with a linear line provides the first estimate of the threshold position.
Note that the data after energy calibration was taken for the step. The threshold position was chosen to be given 
by the position of the half-maximum of the fitting range counts. Final refinements such as including of the quenching 
effects and crass-talk corrections are added in the next step.

\subsection{Crosstalk}
\label{sec:xtalk}
\begin{figure}[!t]
\begin{center}
\includegraphics[width=2.5in]{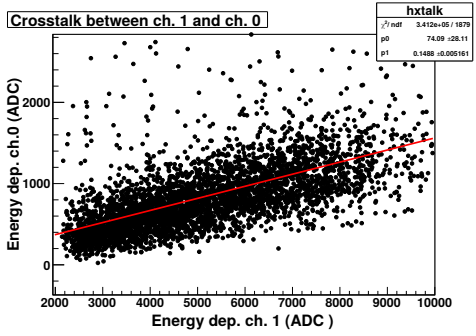}
\caption{An example of energy deposition correlation between two adjacent channels and the fit with a linear line. }
\label{xtalkfit}
\end{center}
\end{figure}
 
Crosstalk phenomenon is an inherent property for the MAPMT based detectors like POLAR.
It caused that a signal induced in a certain channel can be also observed in its neighboring channels.
The crosstalk signals can be attributed either to the multi-channel spread of the scintillating photons in the PMT optical coupling and its entrance glass.
Another reason is a spacial spread of the secondary electrons inside of the PMT.  
The cross-talk modifies the initial energy deposition making the full PMT energy calibration 
and corrections of its non-uniformities more difficult.
An accurate knowledge of the crosstalk is necessary to determine both parameters properly.
Crosstalk factors were obtained by studying the correlation between the recorded energy depositions
of the two channels.
Fig.~\ref{xtalkfit} shows an example of such correlation. The data were taken with the Cs-137 and Na22 sources.
For the analysis one selected events for which the primary channel  (hit by the photon) had the maximum energy deposition. 
The wide band of events shown in the correlation plot is caused by the crosstalk.
The events in the band were fitted with a linear line. The slope of th line gives a crosstalk factor between the primary (hit) 
and the secondary (cross-talk) channel. 
It has been shown that the slopes obtained from the fit are rather weakly dependent in the fit range.
Repeating the same study for all pairs of channels provides a 64$\times$64 matrix, called the crosstalk matrix.
It can be used to correct the crosstalks between channels \cite{silvio, theory}. It was found that the cross-talk 
factors within the same module may reach the values of about ten per cent while the cross-talk between different modules is negligible.
Note that space background data is suitable for the calibration of crosstalk factors as well.

\section{Calibration with mono-energetic polarized beams}

\subsection{Experimental details}
ESRF in France is capable of providing mono-energetic, 100\% polarized and narrow-size X-ray beams in the energy range between 18 keV and 800 keV. 
It provides ideal conditions for energy calibration and modulation factor measurement of POLAR.
In May 2015, we performed a series of calibrations of the POLAR flight model (FM) in the ESRF beam-line ID11.
The beam is 100\% polarized and its size is equal to 0.6  $\times$0.6 mm$^2$. 
The intensity of the beam was reduced to a few thousands of photons/s with an Aluminum wedge in order to avoid event pile-ups.

\begin{figure}[!t]
\begin{center}
\includegraphics[width=2.7in]{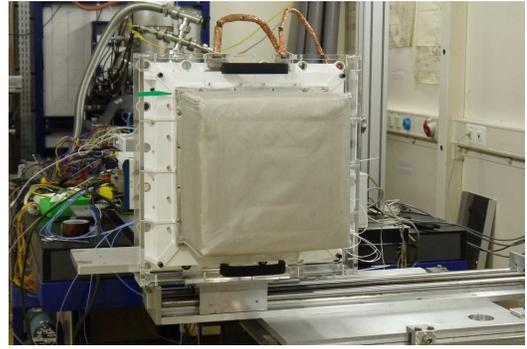}
\caption{A photo of POLAR Flight Model in ESRF beamline ID11. }
\label{esrf2015}
\end{center}
\end{figure}

Fig.~\ref{esrf2015} shows a photo of the FM during tests in the ESRF beam-line ID11. 
The FM was installed on a X-Y table driven by two motors. 
The table allows for directing of the beam into any point of the top surface of the FM via horizontal and vertical translations. 
Before the test, the high voltage and ASIC settings of each module were optimized with the laboratory gamma-rays sources.
During the calibration the FM was tested at four different energies: 140 keV, 110 keV, 80 keV and 60 keV. 
At each beam energy the surface of the FM was scanned with a speed of 5 mm/s at 0$^\circ$, 30$^\circ$ and 60$^\circ$ X-ray incident angles. 
The tests at different incident angles allowed to reproduce a flat illuminations of the instrument expected from the GRBs coming from different directions.
Changing of the incident beam angle was accomplished by rotating the FM along the beam axis manually.

\subsection{Data analysis and results}
\begin{figure}[!t]
\begin{center}
\includegraphics[width=2.7in]{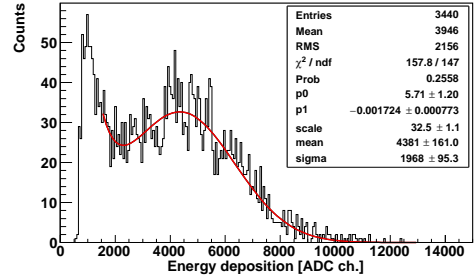}
\centering
\caption{An example of energy spectrum measured with the 80 keV beam. }
\label{fig:efit}
\end{center}
\end{figure}
\begin{figure}[!t]
\begin{center}
\includegraphics[width=2.7in]{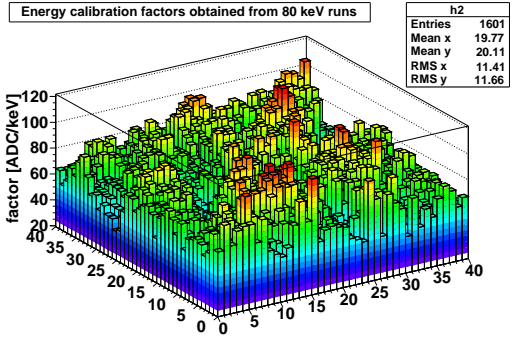}
\caption{Energy calibration factors of POLAR flight model obtained from the ESRF beam test. }
\label{fig:calibrationfactor}
\end{center}
\end{figure}
\begin{figure}[!t]
\begin{center}
\includegraphics[width=2.7in]{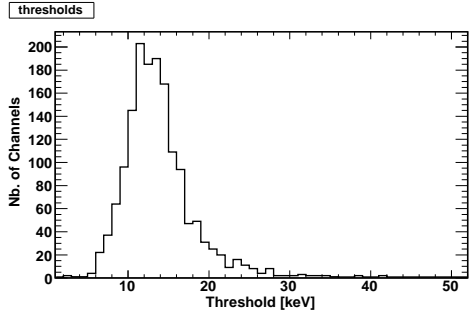}
\centering
\caption{FM threshold distribution at the optimized high voltage and threshold setting. }
\label{thresholddistri}
\end{center}
\end{figure}

The data taken with the 80 keV beam was used for energy calibration. 
Fig.~\ref{fig:efit} shows a typical energy spectrum of one of the POLAR channels.  
Note that only the hits which were produced when the beam was on the corresponding bar 
were selected. The peak in the energy spectrum, according to Monte Carlo simulation,  is due to a full absorption of the 80 keV photons. 
Fitting the peak with a sum of the Gaussian and exponential functions provided the energy calibration factor.
The full distribution of the energy calibration factors for all the channels is shown in Fig.~\ref{fig:calibrationfactor}. 
The same method as described in Section \ref{sec:thr} was used to determine the low energy thresholds.
The distribution of the thresholds is shown for all the channel in Fig.~\ref{thresholddistri}. 
At the optimized settings for both the high voltages and the thresholds, most of the threshold values are between 5 keV and 30 keV .

The crosstalk factors were obtained with the data taken with the 140 keV beam 
using the same method described in Section \ref{sec:xtalk}.
This study provided again the $64\times64$ crosstalk matrix for each detection module . 
As we already shown, the cross-talk studies have been done for two different photon energies. Thus it is possibility to compare and cross-check the measurements. 
The data from the ESRF tests were optimized for a dynamic range of about 200 keV maximum while the data from the laboratory
studies with the Cs-137 source had the dynamic range larger by at least a factor of three. 
Both results are currently evaluated and analyzed with the goal of harmonizing of both sets of data for the whole POLAR energy range. 

Experimental measurements allowed for construction of the POLAR genuine response matrix. The whole response matrix can be based on a set of sub-matrices defined each module as follows \cite{theory}:
\begin{equation}
\vecm{R} =\vecm{F}^{'} \vecm{M},
\label{eq:mres}
\end{equation}
where $\vecm{F}$ and $\vecm{M}$ are the crosstalk matrix and the energy conversion matrix respectively. Note that the energy conversion matrix $\vecm{M}$ is diagonal, whose diagonal elements represent the energy conversion factors of
the corresponding channels.  Both $\vecm{F}$ and $\vecm{M}$   are 64 $\times$ 64 matrices.

The real energy depositions $\vect{Y}=(y_1,y_2,\cdots,y_{63})'$ were reconstructed module by module by performing  linear
transformations:
\begin{equation}
\vect{Y}=\vecm{R}^{-1} \vect{X},
 \label{eq:reconstruction}
\end{equation}
where $\vect{X}=(x_0,x_1,\cdots,x_{63})'$ is the recorded energy depositions.
Applying the reconstruction method as shown above allows for correction of the
gain non-uniformities and crosstalk contributions simultaneously \cite{theory}. 

\begin{figure}[!t]
\begin{center}
\includegraphics[width=1.7in]{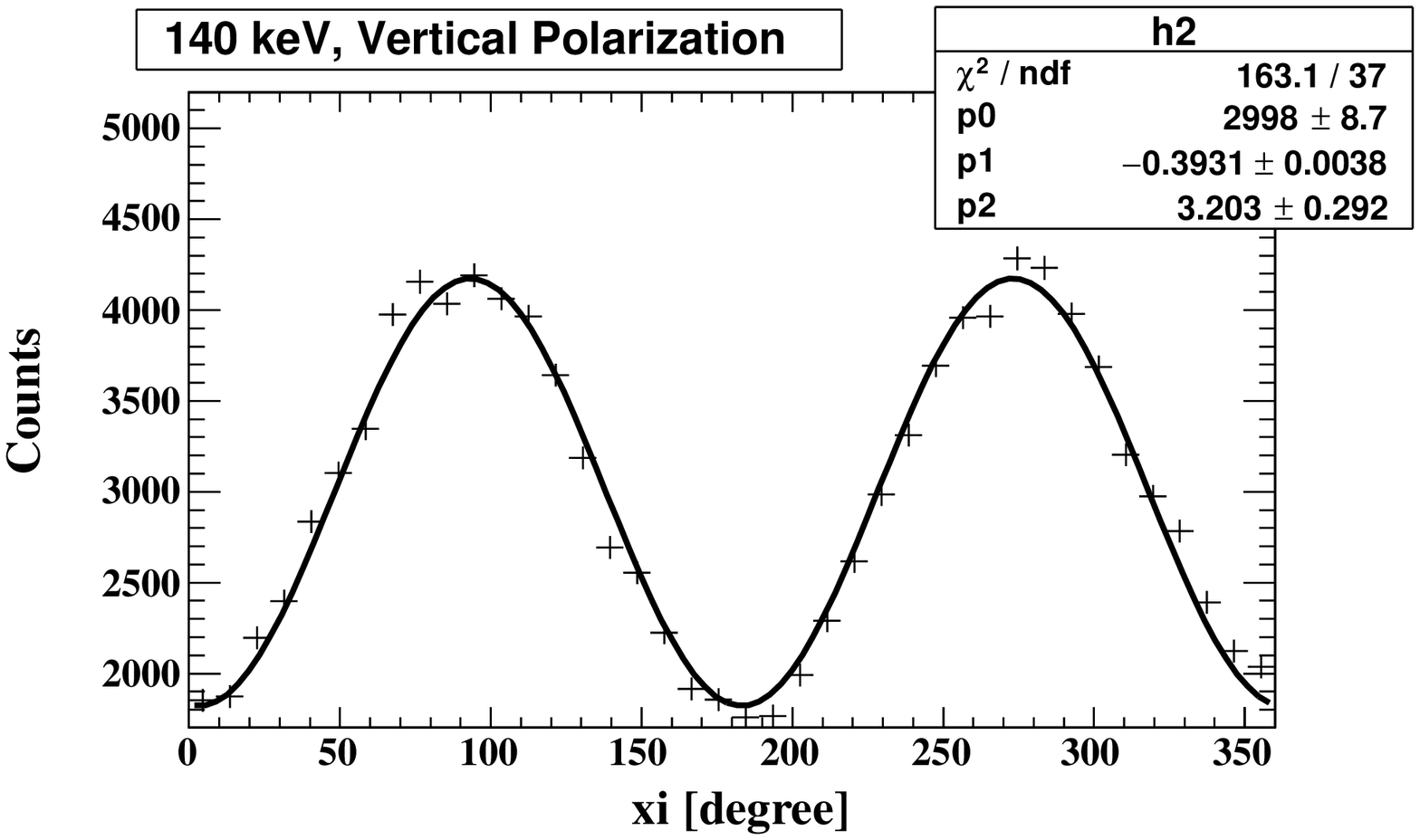}
\includegraphics[width=1.7in]{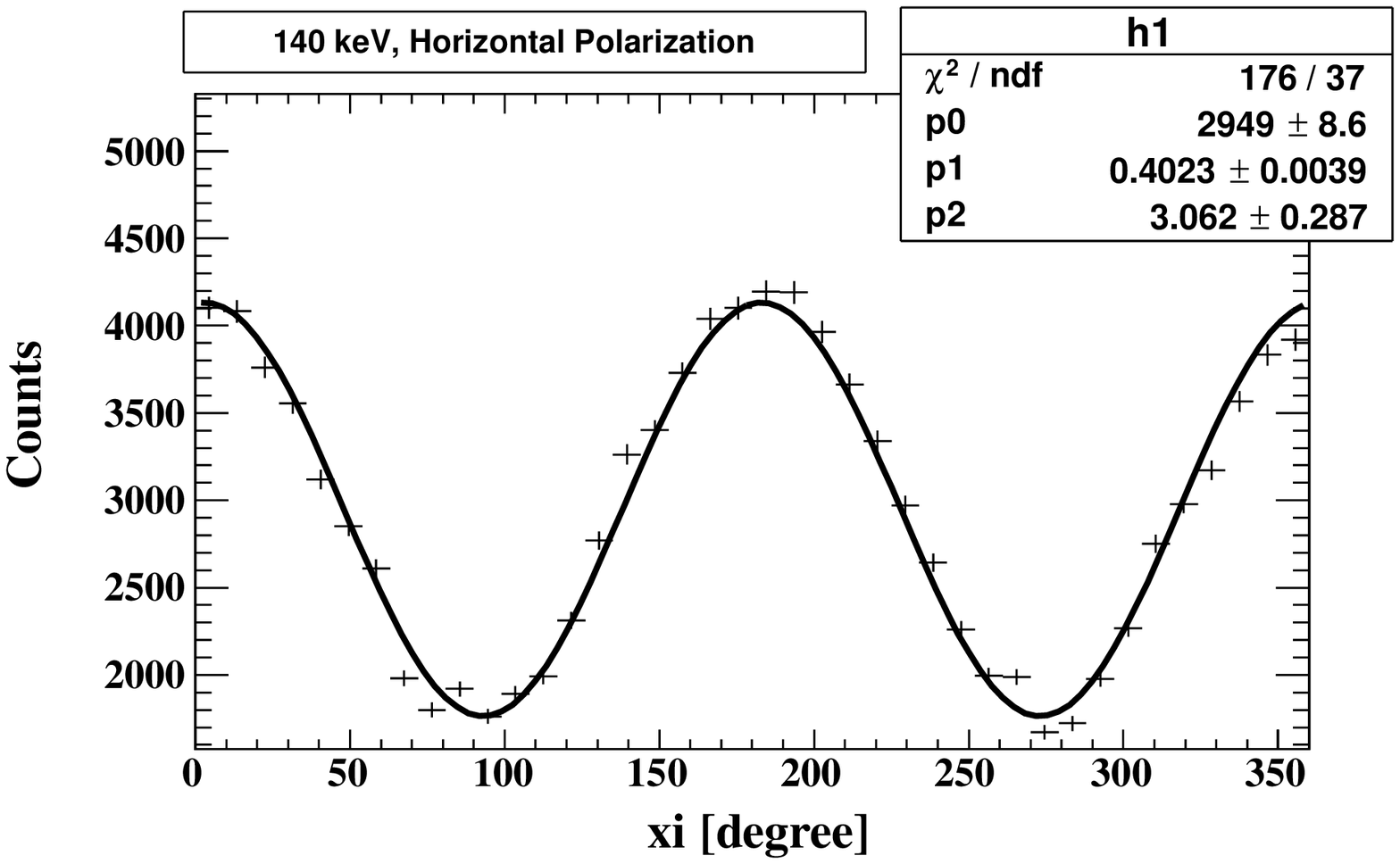}
\caption{Modulation curves measured for the 100\% polarized 140 keV x-ray beam. Left: horizontal polarization. 
Right: vertical polarization. Experimental modulation factors: ~40\%, MC simulation: 40\%. }
\label{fig:modulationcurve}
\end{center}
\end{figure}

The method described in Ref.~\cite{silvio} was used to calculate modulation curves.
The bar with the maximum energy deposition was chosen as the first one that was hit and the bar with the second maximum energy deposition was selected as the second one. 
Two conditions were checked for each event: (1) energy deposition $> 30 $ keV; (2) the two bars were not neighbors.
Due to the limited position resolution of POLAR, the interaction positions in the bar were randomized in order to get smooth distribution of the azimuthal angles from the Compton scattering.
The azimuthal angles were calculated by the formula $\chi =\mathrm{atan}2(y_1-y_2,x_1-x_2)]$.  
The distributions of them builds the modulation curves. 
An example measured for the normal incident angles and X-ray energy of 140 keV is shown in Fig.~\ref{fig:modulationcurve}. 
Note that the geometry effect was corrected from the modulation curves using the method in Ref.~\cite{silvio}.
The preliminary values of the modulation factors for the vertically and horizontally polarized beams are equal to 39.3\% and 40.2\%,
respectively. They are in a very good agreement with Monte Carlo simulations. Further data analysis is ongoing.

\section{Summary and Outlook}
POLAR is a compact space-borne Compton polarimeter devoted to measure
linear polarization in the prompt GRB emission in the energy range from 50 keV to 500 keV. 
The POLAR FM was assembled in January 2015 and is undergoing now further integration tests with the spacecraft. 
POLAR is scheduled for launch onboard the Chinese Spacelab TG-2 in 2016.
In this paper we presented the methods for establishing of the dependence between the energy conversion factor and the  
high voltage value, calibration of the energy thresholds and determination of the crosstalk factors. 
For all above parameters we used a set of laboratory measurements with gamma-ray radioactive sources. 
We also presented the calibration procedure of the POLAR FM used during measurements with the 
polarized X-ray beams at ESRF.
Both the experimental details with techniques used for data analysis as well as some preliminary results are also included . 
For the optimized values of both the gain factors and the threshold values the mean minimum measurable energy of X-rays is around of 12 keV with the FWHM of the distribution equal to about 6 keV. 
Further analysis of the data is still ongoing.
Primary results also show that the modulation factors measured for the 140 keV beam are in good agreement with the Monte Carlo simulations.



\begin{thebibliography}{1}

\bibitem{zb2}
D. Lazzati, ``Polarization in the prompt emission of gamma-ray bursts and their afterglows'', 2006, New J. Phys., vol. 8, pp. 131, 2006.
\bibitem{zb1}K. Toma, T. Sakamoto, B. Zhang, et al., ``Statistical properties of gamma-ray burst polarization'',  Astrophys. J.,  vol. 698, pp. 1042--1053, June 2009.
\bibitem{silvio}S. Orsi, et al., ``Response of the Compton polarimeter POLAR to polarized hard X-rays'',  Nucl. Instr. and Meth. A, vol. 648, pp. 139--154, 2011.
\bibitem{nicolas}N. Produit, et al., ``POLAR, a compact detector for gamma-ray bursts photon
polarization measurements'', Nucl. Instr. and Meth. A, vol. 550, pp. 616--625, 2005.
\bibitem{theory}H.L. Xiao, et al., ``A crosstalk and non-uniformity correction method for the Compact Space-borne Compton Polarimeter
POLAR'', arXiv:1507.04474, 2015.

\end{thebibliography}
%

\end{document}